\documentstyle[12pt,epsfig]{article}

\catcode`@=11
\newcount\@tempcntc
\def\@citex[#1]#2{\if@filesw\immediate\write\@auxout{\string\citation{#2}}\fi
  \@tempcnta\z@\@tempcntb\m@ne\def\@citea{}\@cite{\@for\@citeb:=#2\do
    {\@ifundefined
       {b@\@citeb}{\@citeo\@tempcntb\m@ne\@citea\def\@citea{,}{\bf ?}\@warning
       {Citation `\@citeb' on page \thepage \space undefined}}%
    {\setbox\z@\hbox{\global\@tempcntc0\csname b@\@citeb\endcsname\relax}%
     \ifnum\@tempcntc=\z@ \@citeo\@tempcntb\m@ne
       \@citea\def\@citea{,}\hbox{\csname b@\@citeb\endcsname}%
     \else
      \advance\@tempcntb\@ne
      \ifnum\@tempcntb=\@tempcntc
      \else\advance\@tempcntb\m@ne\@citeo
      \@tempcnta\@tempcntc\@tempcntb\@tempcntc\fi\fi}}\@citeo}{#1}}
\def\@citeo{\ifnum\@tempcnta>\@tempcntb\else\@citea\def\@citea{,}%
  \ifnum\@tempcnta=\@tempcntb\the\@tempcnta\else
   {\advance\@tempcnta\@ne\ifnum\@tempcnta=\@tempcntb \else \def\@citea{--}\fi
    \advance\@tempcnta\m@ne\the\@tempcnta\@citea\the\@tempcntb}\fi\fi}
\catcode`@=12

\begin{document}
\title{\vskip-3cm{\baselineskip14pt
\centerline{\normalsize DESY 00-053\hfill ISSN 0418-9833}
\centerline{\normalsize MPI/PhT/2000-13 \hfill}
\centerline{\normalsize hep-ph/0003297\hfill}
\centerline{\normalsize March 2000\hfill}
}
\vskip1.5cm
Strong Coupling Constant from Scaling Violations in Fragmentation
Functions}
\author{{\sc B.A. Kniehl,$^1$ G. Kramer,$^1$ B. P\"otter$^2$}\\
{\normalsize $^1$ II. Institut f\"ur Theoretische Physik, Universit\"at
Hamburg,}\\
{\normalsize Luruper Chaussee 149, 22761 Hamburg, Germany}\\
{\normalsize $^2$ Max-Planck-Institut f\"ur Physik
(Werner-Heisenberg-Institut),}\\
{\normalsize F\"ohringer Ring 6, 80805 Munich, Germany}}

\date{\today}

\maketitle

\thispagestyle{empty}

\begin{abstract}
We present a new determination of the strong coupling constant $\alpha_s$
through the scaling violations in the fragmentation functions for charged
pions, charged kaons, and protons.
In our fit we include the latest $e^+e^-$ annihilation data from CERN LEP1 and
SLAC SLC on the $Z$-boson resonance and older, yet very precise data from SLAC
PEP at center-of-mass energy $\sqrt s=29$~GeV.
A new world average of $\alpha_s$ is given.

\bigskip

\end{abstract}

\newpage

The strong force acting between hadrons is one of the four fundamental forces
of nature.
It is now commonly believed that the strong interactions are correctly 
described by quantum chromodynamics (QCD), the SU(3) gauge field theory which
contains colored quarks and gluons as elementary particles.
The strong coupling constant $\alpha_s^{(n_f)}(\mu)=g_s^2/(4\pi)$, where $g_s$
is the QCD gauge coupling, is a basic parameter of the standard model of 
elementary particle physics; its value $\alpha_s^{(5)}(M_Z)$ at the $Z$-boson 
mass scale is listed among the constants of nature in the Review of Particle
Physics \cite{pdg}.
Here, $\mu$ is the renormalization scale, and $n_f$ is the number of active 
quark flavors, with mass $m_q\ll\mu$.
The formulation of $\alpha_s^{(n_f)}(\mu)$ in the modified minimal-subtraction
($\overline{\mathrm{MS}}$) scheme, with four-loop evolution and three-loop
matching at the flavor thresholds, is described in Ref.~\cite{cks}.
 
There are a number of processes in which $\alpha_s^{(5)}(M_Z)$ can be measured
(see Refs.~\cite{pdg,alphas}, for recent reviews).
A reliable method to determine $\alpha_s^{(5)}(M_Z)$ is through the extraction
of the fragmentation functions (FF's) in the annihilation process 
\begin{equation}
e^+e^-\to(\gamma,Z)\to h+X,
\label{process}
\end{equation}
which describes the inclusive production of a single charged hadron, $h$.
Here, $h$ may either refer to a specific charged-hadron species, such as
$\pi^\pm$, $K^\pm$, or $p/\bar p$, or to the sum of all charged hadrons.
The partonic cross sections pertinent to process~(\ref{process}) can
entirely be calculated in perturbative QCD with no additional input, except
for $\alpha_s$.
They are known at next-to-leading order (NLO) \cite{3} and even at
next-to-next-to-leading order \cite{rn96}.
The subsequent transition of the partons into hadrons takes place at an energy
scale of the order of 1~GeV and can, therefore, not be treated in perturbation
theory.
Instead, the hadronization of the partons is described by FF's $D_a^h(x,Q^2)$.
Their values correspond to the probability that the parton $a$, which is
produced at short distance, of order $1/Q$, fragments into the hadron $h$
carrying the fraction $x$ of the momentum of $a$.
In the case of process~(\ref{process}), $Q$ is typically of the order of the
center-of-mass (CM) energy $\sqrt s$.
Given their $x$ dependence at some scale $Q_0$, the evolution of the FF's with
$Q$ may be computed perturbatively from the timelike Altarelli-Parisi
equations \cite{ap1}, which are presently known through NLO \cite{ap2}.
This method to determine $\alpha_s^{(5)}(M_Z)$ is particularly clean in the
sense that, unlike other methods, it is not plagued by uncertainties
associated with hadronization corrections, jet algorithms, parton density
functions (PDF's), {\it etc.}
We recall that, similarly to the scaling violations in the PDF's, perturbative
QCD only predicts the $Q^2$ dependence of the FF's.
Therefore, measurements at different CM energies are needed in order to
extract values of $\alpha_s^{(5)}(M_Z)$.
Furthermore, since the $Q^2$ evolution mixes the quark and gluon FF's, it is
essential to determine all FF's individually.

In 1994/95, two of us, together with Binnewies, extracted $\pi^{\pm}$ and
$K^{\pm}$ FF's through fits to PEP and partially preliminary LEP1 data and
thus determined $\alpha_s^{(5)}(M_Z)$ to be 0.118 (0.122) at NLO (LO)
\cite{5} (BKK).
However, these analyses suffered from the lack of specific data on the
fragmentation of tagged quarks and gluons to $\pi^\pm$, $K^\pm$, and
$p/\bar p$ hadrons.
This drawback has been cured in 1998 by the advent of a wealth of new data
from the LEP1 and SLC experiments \cite{9,D,S,gA,gO}. 
The data partly come as light-, $c$-, and $b$-quark-enriched samples with
identified final-state hadrons ($\pi^\pm$,$K^\pm$, and $p/\bar p$) 
\cite{9,D,S} or as gluon-tagged three-jet samples without hadron
identification \cite{gA,gO}.
This new situation motivates us to update, refine, and extend the BKK analysis 
\cite{5} by generating new LO and NLO sets of $\pi^\pm$, $K^\pm$, and
$p/\bar p$ FF's.
By also including in our fits $\pi^\pm$, $K^\pm$, and $p/\bar p$ data (without
flavor separation) from PEP \cite{T}, with CM energy  $\sqrt s=29$~GeV, we
obtain a handle on the scaling violations in the FF's, which allows us to
extract LO and NLO values of $\alpha_s^{(5)}(M_Z)$.
The latter data \cite{T} combines small statistical errors with fine binning
in $x$ and is more constraining than other data from the pre-LEP1/SLC era.

The NLO formalism for extracting FF's from $e^+e^-$ data was comprehensively
described in Ref.~\cite{5}.
We work in the $\overline{\mathrm{MS}}$ renormalization and factorization
scheme and choose the renormalization scale $\mu$ and the factorization scale
$M_f$ to be $\mu=M_f=\xi\sqrt s$, except for gluon-tagged three-jet events,
where we put $\mu=M_f=2\xi E_{\mathrm{jet}}$, with $E_{\mathrm{jet}}$ being
the gluon jet energy in the CM frame.
Here, the dimensionless parameter $\xi$ is introduced to determine the
theoretical uncertainty in $\alpha_s^{(5)}(M_Z)$ from scale variations.
As usual, we allow for variations between $\xi=1/2$ and 2 around the default
value 1.
For the actual fitting procedure, we use $x$ bins in the interval
$0.1\le x\le 1$
and integrate the theoretical functions over the bin widths as is done in the
experimental analyses.
The restriction at small $x$ is introduced to exclude events in the
nonperturbative region, where mass effects and nonperturbative
intrinsic-transverse-momentum effects are important and the underlying
formalism is insufficient.
We parameterize the $x$ dependence of the FF's at the starting scale $Q_0$ as  
\begin{equation}
D_a^h(x,Q_0^2)=Nx^{\alpha}(1-x)^{\beta}.
\label{ansatz} 
\end{equation}
We treat $N$, $\alpha$, and $\beta$ as independent fit parameters.
In addition, we take the asymptotic scale parameter
$\Lambda_{\overline{\mathrm{MS}}}^{(5)}$, appropriate for five quark flavors,
as a free parameter.
Thus, we have a total of 46 independent fit parameters.
The quality of the fit is measured in terms of the $\chi^2$ value per degree
of freedom, $\chi^2_{\mathrm{DF}}$, for all selected data points.
Using a multidimensional minimization algorithm \cite{18}, we search this
46-dimensional parameter space for the point at which the deviation of the
theoretical prediction from the data becomes minimal.

\begin{table}[hhh]
\caption{CM energies, types of data, and $\chi^2_{\mathrm{DF}}$ values
obtained at LO and NLO for the various data samples. \vspace{0mm}}
\label{Table1}
\begin{center}
\begin{tabular}{c|l|ll} \hline \hline 
 $\sqrt{s}$ [GeV] & Data type & 
 \multicolumn{2}{c}{\makebox[4.5cm][c]{$\chi^2_{\mathrm{DF}}$ in NLO (LO)}} \\
 \hline
29.0 & $\sigma^\pi$~(all) & 0.64 (0.71) \cite{T} & \\
     & $\sigma^K$~(all) & 1.86 (1.40) \cite{T} & \\
     & $\sigma^p$~(all) & 0.79 (0.70) \cite{T} & \\ \hline
91.2 & $\sigma^h$~(all) & 1.28 (1.40) \cite{D} & 1.32 (1.44) \cite{S} \\
     & $\sigma^h$~(uds) & 0.20 (0.20) \cite{D} & \\
     & $\sigma^h$~(b)   & 0.43 (0.41) \cite{D} & \\ \hline
     & $\sigma^\pi$~(all) & 1.28 (1.65) \cite{9} & \\
     &                    & 0.58 (0.60) \cite{D} & 3.09 (3.13) \cite{S} \\
     & $\sigma^\pi$~(uds) & 0.72 (0.73) \cite{D} & 1.87 (2.17) \cite{S} \\
     & $\sigma^\pi$~(c) & & 1.36 (1.16) \cite{S} \\
     & $\sigma^\pi$~(b) & 0.57 (0.58) \cite{D} & 1.00 (0.99) \cite{S} \\ \hline
     & $\sigma^K$~(all) & 0.30 (0.32) \cite{9} & \\
     &                  & 0.86 (0.79) \cite{D} & 0.44 (0.45) \cite{S} \\
     & $\sigma^K$~(uds) & 0.53 (0.60) \cite{D} & 0.65 (0.64) \cite{S} \\
     & $\sigma^K$~(c) & & 2.11 (1.90) \cite{S} \\
     & $\sigma^K$~(b) & 0.14 (0.14) \cite{D} & 1.21 (1.23) \cite{S} \\ \hline
     & $\sigma^p$~(all) & 0.93 (0.80) \cite{9} & \\
     &                  & 0.09 (0.06) \cite{D} & 0.79 (0.70) \cite{S} \\
     & $\sigma^p$~(uds) & 0.11 (0.14) \cite{D} & 1.29 (1.28) \cite{S} \\
     & $\sigma^p$~(c)   & & 0.92 (0.89) \cite{S} \\
     & $\sigma^p$~(b)   & 0.56 (0.62) \cite{D} & 0.97 (0.89) \cite{S} \\ \hline
$E_{\mathrm{jet}}$ [GeV] & & & \\ \hline
26.2 & $D_g^h$ & 1.19 (1.18) \cite{gA} & \\
40.1 & $D_g^h$ & 1.03 (0.90) \cite{gO} & \\ \hline \hline
\end{tabular}
\end{center}
\end{table}

\begin{figure}[hhh]
  \unitlength1mm
  \begin{picture}(122,160)
    \put(3,0){\epsfig{file=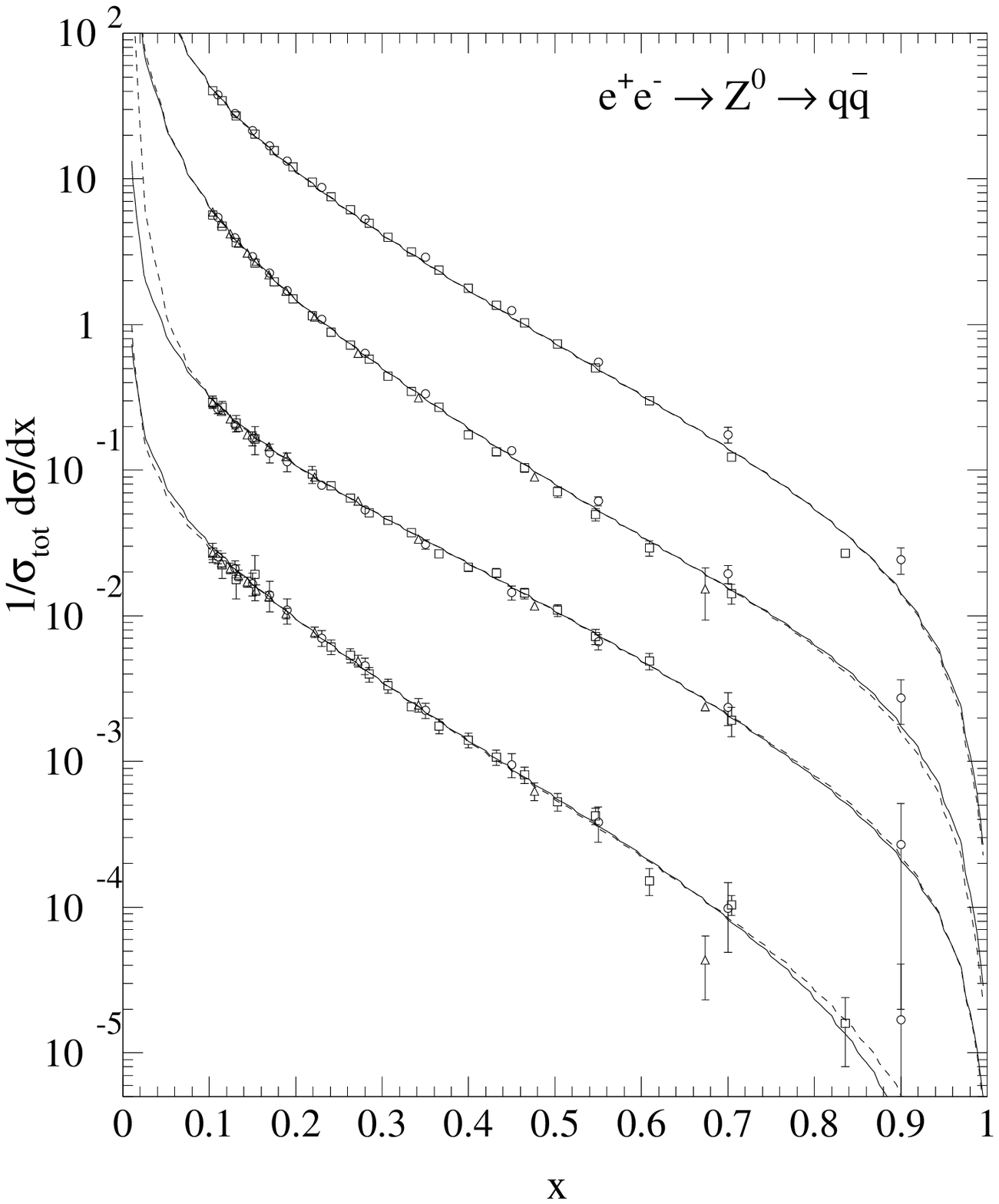,width=16cm}}
  \end{picture}
\caption{Normalized differential cross section of inclusive hadron production
at $\protect\sqrt{s}=91.2$~GeV as a function of $x$.
The LO (dashed lines) and NLO (solid lines) fit results are compared with data
from ALEPH \protect\cite{9} (triangles), DELPHI \protect\cite{D} (circles),
and SLD \protect\cite{S} (squares).
The upmost, second, third, and lowest curves refer to charged hadrons,
$\pi^\pm$, $K^\pm$, and $p/\bar{p}$, respectively.
Each pair of curves is rescaled relative to the nearest upper one by a factor
of 1/5.}
\label{Figure1}
\end{figure}

\begin{figure}[hhh]
  \unitlength1mm
  \begin{picture}(122,160)
    \put(3,0){\epsfig{file=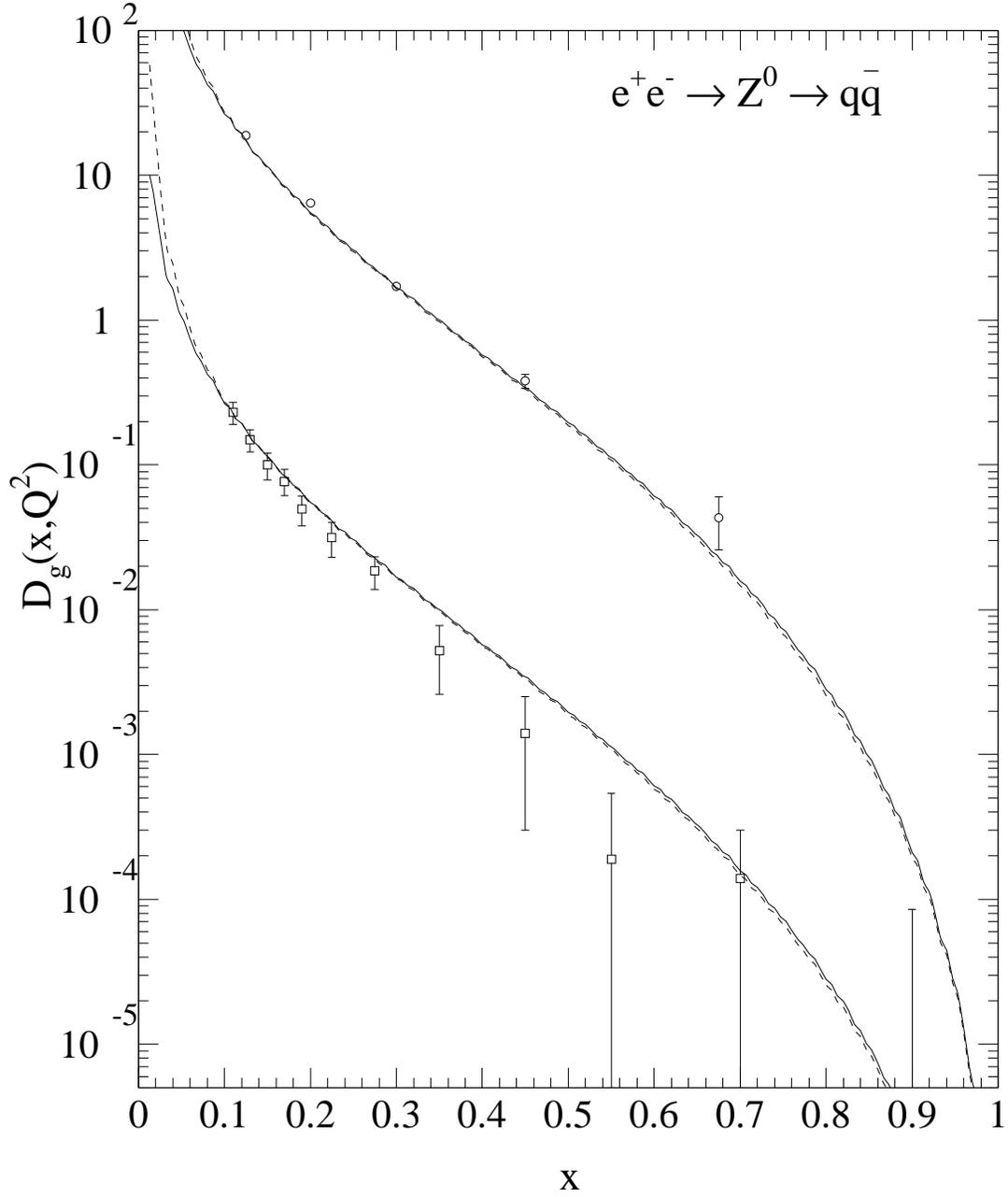,width=16cm}}
  \end{picture}
\caption{Gluon FF for charged-hadron production at $Q=52.4$ and 80.2~GeV as a
function of $x$.
The LO (dashed lines) and NLO (solid lines) predictions are compared with 
three-jet data from ALEPH \protect\cite{gA} with $E_{\mathrm{jet}}=26.2$~GeV
(upper curves) and from OPAL \protect\cite{gO} with
$E_{\mathrm{jet}}=40.1$~GeV (lower curves).
The OPAL data and the pertinent predictions are rescaled by a factor of
1/100.}  
\label{Figure2}
\end{figure}

The $\chi^2_{\mathrm{DF}}$ values achieved for the various data sets used in
our LO and NLO fits may be seen from Table~\ref{Table1}.
Most of the $\chi^2_{\mathrm{DF}}$ values lie around unity or below,
indicating that the fitted FF's describe all data sets within their respective
errors. In general, the $\chi^2_{\mathrm{DF}}$ values come out
slightly in favor for the DELPHI \cite{D} data. The overall goodness
of the NLO (LO) fit is given by $\chi^2_{\mathrm{DF}}=0.98$ (0.97).
The goodness of our fit may also be judged from Figs.~\ref{Figure1} and
\ref{Figure2}, where our LO and NLO fit results are compared with the ALEPH
\cite{9,gA}, DELPHI \cite{D}, OPAL \cite{gO}, and SLD \cite{S} data.
In Fig.~\ref{Figure1}, we study the differential cross section
$(1/\sigma_{\mathrm{tot}})d\sigma^h/dx$ of process~(\ref{process}) for 
$\pi^\pm$, $K^\pm$, $p/\bar p$, and unidentified charged hadrons at
$\sqrt{s}=91.2$~GeV, normalized to the total hadronic cross section
$\sigma_{\mathrm{tot}}$, as a function of the scaled momentum
$x=2p_h/\sqrt s$. As in Refs.~\cite{D,S}, we assume that the sum of
the $\pi^\pm$, $K^\pm$, and $p/\bar p$ data exhaust the full
charged-hadron data. We observe that, in all cases, the various data
are mutually consistent with each other and are nicely described by
the LO and NLO fits, which is also reflected in the relatively small
$\chi^2_{\mathrm{DF}}$ values given in  Table~\ref{Table1}.
The LO and NLO fits are almost indistinguishable in those regions of $x$,
where the data have small errors. At large $x$, where the statistical
errors are large, the LO and NLO results  sometimes moderately deviate
from each other. In Fig.~\ref{Figure2}, we compare the ALEPH \cite{gA}
and OPAL \cite{gO} measurements of the gluon FF in gluon-tagged
charged-hadron production, with $E_{\mathrm{jet}}=26.2$ and 40.1~GeV,
respectively, with our LO and NLO fit results. The data are nicely
fitted, with $\chi^2_{\mathrm{DF}}$ values of order unity, as may be
seen from Table~\ref{Table1}. By the same token, this implies that the
data \cite{gA,gO} are mutually consistent\footnote{The new FF sets
can be obtained from http://www.desy.de/\~{}poetter/kkp.html}.

The purpose of this letter is to update and improve the determinations of
$\Lambda_{\overline{\mathrm{MS}}}^{(5)}$ and $\alpha_s^{(5)}(M_Z)$ from the
scaling violations in the FF's.
We obtain
$\Lambda_{\overline{\mathrm{MS}}}^{(5)}=88{+34\atop-31}{+3\atop-23}$~MeV at
LO and
$\Lambda_{\overline{\mathrm{MS}}}^{(5)}=213{+75\atop-73}{+22\atop-29}$~MeV at
NLO, where the first errors are experimental and the second ones are
theoretical.
The experimental errors are determined by varying
$\Lambda_{\overline{\mathrm{MS}}}^{(5)}$ in such a way that the total
$\chi^2_{\mathrm{DF}}$ value is increased by one unit if all the other fit
parameters are kept fixed, while the theoretical errors are obtained by
repeating the LO and NLO fits for the scale choices $\xi=1/2$ and 2.
From the LO and NLO formulas for $\alpha_s^{(n_f)}(\mu)$ \cite{cks}, we thus
obtain
\begin{eqnarray}
\alpha_s^{(5)}(M_Z)&=&0.1181{+0.0058\atop-0.0069}{+0.0006\atop-0.0049}\qquad
\mbox{(LO)},\nonumber\\
\alpha_s^{(5)}(M_Z)&=&0.1170{+0.0055\atop-0.0069}{+0.0017\atop-0.0025}\qquad
\mbox{(NLO)},
\label{as}
\end{eqnarray}
respectively.
Adding the maximum experimental and theoretical deviations from the central 
values in quadrature, we find
$\Lambda_{\overline{\mathrm{MS}}}^{(5)}=(88\pm41)$~MeV and
$\alpha_s^{(5)}(M_Z)=0.1181\pm0.0085$ at LO and
$\Lambda_{\overline{\mathrm{MS}}}^{(5)}=(213\pm79)$~MeV and
$\alpha_s^{(5)}(M_Z)=0.1170\pm0.0073$ at NLO.
We observe that our LO and NLO values of $\alpha_s^{(5)}(M_Z)$ are quite
consistent with each other, which indicates that our analysis is 
perturbatively stable.
The fact that the respective values of
$\Lambda_{\overline{\mathrm{MS}}}^{(5)}$ significantly differ is a well-known
feature of the $\overline{\mathrm{MS}}$ definition of
$\alpha_s^{(n_f)}(\mu)$ \cite{cks}.

\begin{figure}[hhh]
  \unitlength1mm
  \begin{picture}(122,160)
    \put(3,0){\epsfig{file=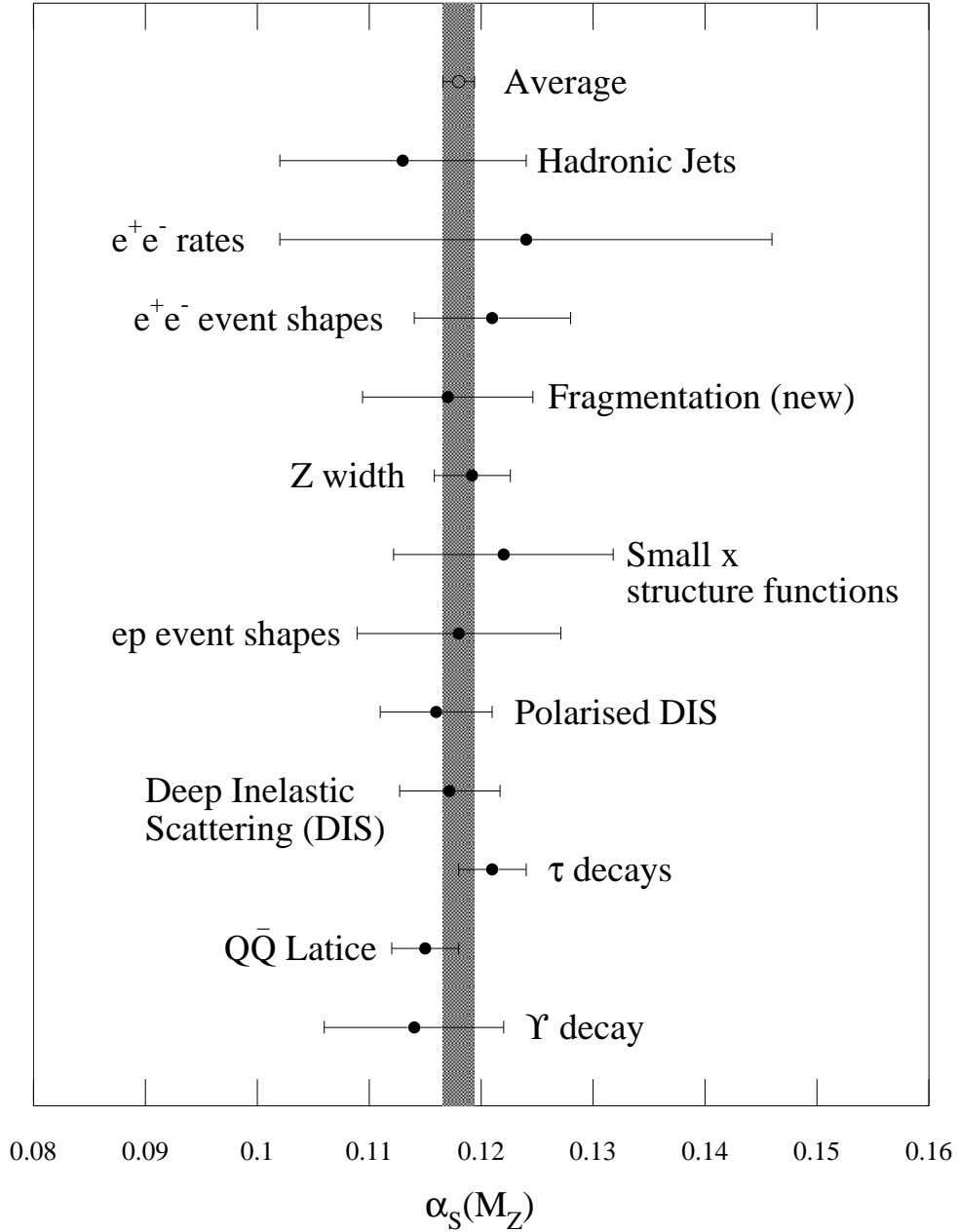,width=16cm}}
  \end{picture}
\caption{Summary of the values of $\alpha_s^{(5)}(M_Z)$ from various
processes.
The errors shown represent the total errors including theoretical
uncertainties.}  
\label{Figure3}
\end{figure}

Our values of $\Lambda_{\overline{\mathrm{MS}}}^{(5)}$ and
$\alpha_s^{(5)}(M_Z)$ perfectly agree with those presently quoted by the
Particle Data Group (PDG) \cite{pdg} as world averages,
$\Lambda_{\overline{\mathrm{MS}}}^{(5)}=212{+25\atop-23}$~MeV and
$\alpha_s^{(5)}(M_Z)=0.1185\pm0.002$, respectively.
Notice that, in contrast to our LO and NLO analyses, the PDG evaluates
$\Lambda_{\overline{\mathrm{MS}}}^{(5)}$ from $\alpha_s^{(5)}(M_Z)$ using the
three-loop relationship \cite{cks}.
The PDG combines twelve different kinds of $\alpha_s^{(5)}(M_Z)$ measurements, 
including one from the scaling violations in the FF's \cite{bus}, by
minimizing the total $\chi^2$ value.
The world average cited above is then estimated from the outcome by allowing
for correlations between certain systematic errors.
It is interesting to investigate how the world average of
$\alpha_s^{(5)}(M_Z)$ is affected by our analysis.
To this end, we first combine the twelve $\alpha_s^{(5)}(M_Z)$ measurements
reported in Ref.~\cite{pdg} to find $\alpha_s^{(5)}(M_Z)=0.1181\pm0.0014$ with
$\chi^2=3.74$.\footnote{This result slightly differs from the corresponding 
one found in Ref.~\cite{pdg}.}
If we replace the value $\alpha_s^{(5)}(M_Z)=0.125\pm0.005\pm0.008$ resulting
from previous FF analyses \cite{bus}, which enters the PDG average, with our
new NLO value, then we obtain $\alpha_s^{(5)}(M_Z)=0.1180\pm0.0014$ with
$\chi^2=3.21$, {\it i.e.}, the face value of the world average essentially
goes unchanged, while the overall agreement is appreciably improved.
This is also evident from the comparison of Fig.~\ref{Figure3}, which
summarizes our updated world average, with the corresponding Fig.~9.1 in
Ref.~\cite{pdg}.
We observe that the central value of our new NLO result for
$\alpha_s^{(5)}(M_Z)$ falls into the shaded band, which indicates the error of
the world average, while in Fig.~9.1 of Ref.~\cite{pdg} the corresponding 
central value \cite{bus} exceeds the world average by 3.3 standard deviations
of the latter, which is more than for all other eleven processes.
Furthermore, our new NLO result has a somewhat smaller error (0.0073) than the
corresponding result \cite{bus} used by the PDG (0.009).
If we take the point of view that our new NLO value of $\alpha_s^{(5)}(M_Z)$
should rather be combined with the result from the previous FF analyses 
\cite{bus} before taking the world average, then the latter turns out to be
$\alpha_s^{(5)}(M_Z)=0.1181\pm0.0014$ with $\chi^2=3.29$.

In summary, we presented an updated and improved determination of
$\alpha_s^{(5)}(M_Z)$ from the LO and NLO analyses of inclusive light-hadron
production in $e^+e^-$ annihilation.
Our strategy was to only include in our fits high-precision LEP1 and SLC data
with both flavor separation and hadron identification (namely, light-, $c$-,
and $b$-quark-enriched samples of $\pi^\pm$, $K^\pm$, and $p\bar p$ data)
\cite{S,D}, gluon-tagged three-jet samples with a fixed gluon-jet energy
\cite{gA,gO}, and the $\pi^\pm$, $K^\pm$, and $p/\bar p$ data sets from the
pre-LEP1/SLC era with the highest statistics and the finest binning in $x$
\cite{T}.
Our LO and NLO results for $\alpha_s^{(5)}(M_Z)$ are given in Eq.~(\ref{as}).
They should be compared with the result from scaling violations in FF's quoted
in Ref.~\cite{pdg}, $0.125\pm0.005\pm0.008$ \cite{bus}.
If we repeat the global analysis of Ref.~\cite{pdg} with this result replaced
by our new NLO value, then the world average (before taking into account
estimated correlations between systematic errors) is changed from
$\alpha_s^{(5)}(M_Z)=0.1181\pm0.0014$ with $\chi^2=3.74$ to
$\alpha_s^{(5)}(M_Z)=0.1180\pm0.0014$ with $\chi^2=3.21$, {\it i.e.}, the
overall agreement is appreciably improved.

The II. Institut f\"ur Theoretische Physik is supported by the
Bundesministerium f\"ur Bildung und Forschung under Contract No.\ 05~HT9GUA~3,
and by the European Commission through the Research Training Network
{\it Quantum Chromodynamics and the Deep Structure of Elementary Particles}
under Contract No.\ ERBFMRXCT980194.

\newpage

\newpage

\newpage

\end{document}